\documentclass[a4paper,11pt]{article}

\usepackage{jheppub}
\usepackage{amsmath,amssymb,amscd}
\usepackage{bbm}
\usepackage{graphicx,xcolor}
\usepackage{url}
\usepackage[export]{adjustbox}
\usepackage{accents}
\usepackage[]{algorithm2e}
\usepackage{verbatim}

\graphicspath{ {fig/}{../fig} }

\makeatletter
\def\hlinewd#1{%
  \noalign{\ifnum0=`}\fi\hrule \@height #1 \futurelet
   \reserved@a\@xhline}
\makeatother

\makeatletter
\renewcommand\@fpheader{}
\renewcommand\@journal{}
\makeatother

\definecolor{darkgreen}{rgb}{0.,.3,0}
\definecolor{darkblue}{rgb}{0.0,0.0,0.5}

\newcommand{\Reduze}{\texttt{Reduze\;2}}
\newcommand{\ud}{\mathrm{d}}

\newcommand{\hide}[1]{}

\title{
{Planar two-loop integrals for $\mathbf{\mu e}$ scattering in QED with finite lepton masses
}}
\preprint{MITP/21-024}
\author[a]{Matthias Heller}

\affiliation[a]{Institut f\"ur Kernphysik und PRISMA$^+$ Cluster of Excellence, Johannes-Gutenberg Universit\"{a}t,\\
55099 Mainz, Deutschland}

\emailAdd{maheller@students.uni-mainz.de}

\abstract{We present analytic results for one of two types of planar QED two-loop integrals for $\mu e$ scattering including finite lepton masses. No approximations are made, such that the results are valid not only in the limit of a small electron mass. The results are expressed in terms of multiple polylogarithms with algebraic function arguments in a representation which allows for fast numerical evaluation in the physical phase-space.
}

\begin{document}
\maketitle

\allowdisplaybreaks[4]

\section{Introduction}
Recent results of the $g-2$ experiment at Fermi Lab have drawn a lot of attention in the community showing clear sign of new physics. The reported discrepancy between the measured and theoretical value of the anomalous magnetic moment of the muon is now, combining the new measurement with old results from the Brookhaven experiment, approximately $4.2$ standard deviations \cite{Abi:2021gix}. The main limitation from the theory side are stemming from strong interactions, which are calculated from dispersion theory. To decrease the theoretical uncertainties, a new experiment, MUonE, was proposed at Cern. The main objective of this experiment is to determine the leading order hadronic contribution to the anomalous magnetic moment of the muon by measuring the effective electromagnetic coupling constant $\alpha$ \cite{Abbiendi:2016xup}. For a better theoretical understanding of the $\mu e$ scattering experiment, a precise knowledge of the size of radiative corrections is necessary. In this work, we give one important ingredient to deepen that understanding: we calculate one part of the contributing Feynman integrals, that need to be known in order to calculate the two-loop QED corrections to that process. Up to now, those integrals are only known in the limit of a vanishing electron mass \cite{DiVita:2018nnh,Mastrolia:2017pfy} and in the equal-mass limit of Bhabha scattering \cite{Henn:2013woa}. We express our integrals in terms of multiple polylogarithms. These function admit a fast numerical evaluation. For instance in Ref. \cite{Heller:2019gkq}, the numerical evaluation of all master integrals takes approximately $0.5$ seconds for a generic phase space point.

The calculation presented in this work is also interesting from a mathematical point of view. In the case of interest it is possible to derive a differential equation in canonical form, i.e. a differential equation in which the dimensional regulator decouples from the kinematical part and whose functional behavior with respect to external scales can be written in d-log form \cite{Henn:2013pwa,Kotikov:2010gf}. However, non-rationalizable algebraic function arguments appear inside those d-logs, which renders the integration difficult. Although it has become clear recently that not all differential equations in d-log form admit a solution in terms of multiple polylogarithms \cite{Brown:2020rda}, we find that in the present case such a solution can be found. This generalizes the findings of Ref. \cite{Heller:2019gkq}, where it was shown that through direct integration in the equal-mass case the only integral that was not solved in terms of generalized polylogarithms in Ref. \cite{Henn:2013woa} could also be expressed in terms of these functions. After the recent success in finding a polylogarithmic solutions to the two-loop integrals for mixed QCD-EW corrections to the Drell-Yan process \cite{Heller:2019gkq,Heller:2020owb}, this is to our knowledge the second time in which Feynman integrals with unrationalizable square roots inside the symbol could be matched to an ansatz of multiple polylogarithms and the first time in which this has been done with several square roots. This raises the question if Feynman integrals have additional properties, which allows to integrate them always in terms of multiple polylogarithms, once a d-log form of their differential equation is found.

The article is organized as follows. In Sec. \ref{sec:kinematics} we define the kinematics for $\mu e$ scattering and parametrize the physical phase-space. In Sec. \ref{sec:basis} we introduce the integral family considered in this work and define a canonical basis for the master integrals. In Sec. \ref{sec:deq_simple} we show how to simplify the canonical differential equation such that we are able to match its solution. In Sec. \ref{sec:integration} we review how to find an ansatz that matches the symbol of the differential equation. In Sec. \ref{sec:checks}, we present our results and comment on the numerical evaluation. We give an outlook in Sec. \ref{sec:outlook}.

\section{Kinematics}
\label{sec:kinematics}
We study a two-to-two scattering process of electrons and muons with momenta $p_1,\dots p_4$,
\begin{equation}
    \mu(p_1)+e(p_2)\rightarrow \mu(p_3)+e(p_4),
\end{equation}
where $p_1^2=p_3^2=m_1^2$ and $p_2^2=p_4^2=m_2^2$. In the following we assume without loss of generality $m_2\leq m_1$. As usual, we define Mandelstam variables as
\begin{equation}
    s=(p_1+p_2)^2,\quad t=(p_1-p_3)^2,\quad u=(p_1-p_4)^2.
\end{equation}
Due to the on-shell relation
\begin{equation}
    s+t+u=2m_1^2+2m_2^2,
\end{equation}
we only use $s$ and $t$ in the following.

Next we need to derive the limits for $s$ and $t$ in the physical phase-space scattering region. Clearly for the center-of-mass energy we have
\begin{equation}
    s>(m_1+m_2)^2.
\end{equation}
In the center-of-mass frame we can parametrize the incoming momenta as follows,
\begin{equation}
    p_1=\left(
\begin{array}{c}
E_1\\
0\\
0\\
p
\end{array}
\right),\quad     p_2=\left(
\begin{array}{c}
E_2\\
0\\
0\\
-p
\end{array}
\right),
\end{equation}
where
\begin{equation}
    E_1=\frac{m_1^2-m_2^2+s}{2\sqrt{s}},\quad E_2=\frac{m_2^2-m_1^2+s}{2\sqrt{s}},\qquad p=\frac{\sqrt{m_1^4+(m_2^2-s)^2-2m_1^2(m_2^2+s)}}{2\sqrt{s}}.
\end{equation}
Denoting the scattering angle in the center-of-mass frame by $\theta$ it follows that the final state momenta are given by
\begin{equation}
    p_3=\left(
\begin{array}{c}
E_1\\
p\sin(\theta)\\
0\\
p\cos(\theta)
\end{array}
\right),\quad     p_4=\left(
\begin{array}{c}
E_2\\
-p\sin(\theta)\\
0\\
-p\cos(\theta)
\end{array}
\right).
\end{equation}
We can now derive the minimum and maximum allowed values for $t$ by considering forward ($\theta=0$) and backward ($\theta=\pi$) scattering:
\begin{equation}
    t_{\mathrm{min}}=-\frac{m1^4+(m_2^2-s)^2-2m_1^2(m_2^2+s)}{s},\quad t_{\mathrm{max}}=0.
\end{equation}
Therefore, the physical phase-space of the scattering process considered here, is given by
\begin{equation}
    s>(m_1+m_2)^2,\qquad -\frac{m1^4+(m_2^2-s)^2-2m_1^2(m_2^2+s)}{s}<t<0.
\end{equation}

In what follows we use the dimensional quantities
\begin{equation}
    x=\frac{m_1^2+m_2^2}{2s},\qquad y=\frac{m_1^2-m_2^2}{2s},\qquad z=\frac{t}{s}.
\end{equation}
In terms of $x$, $y$ and $z$, the physical phase-space is parametrized by
\begin{equation}
    0<y<\frac{1}{2},\quad y<x<\frac{1}{4}\big(1+4y^2\big),\quad -1+4x-4y^2<z<0.
\end{equation}

\section{A canonical basis for the master integrals}
\label{sec:basis}

We define the integral family $\mathrm{I}_{n_1,\dots,n_9}$ with the following set of $9$ propagators:                
\begin{align}
&D_1=k_1^2-m_1^2,&
&D_2=k_2^2-m_1^2,&
&D_3=(k_1-k_2)^2,&\nonumber\\
&D_4=(k_1-p_1)^2,&
&D_5=(k_2-p_1)^2,&
&D_6=(k_1-p_1-p_2)^2-m_2^2,&\nonumber\\
&D_7=(k_2-p_1-p_2)^2-m_2^2,&
&D_8=(k_1-p_3)^2,&
&D_9=(k_2-p_3)^2&.
\end{align}
For a scalar integral from this family we employ the integration measure
\begin{equation}
    \mathrm{I}_{n_1,\dots,n_9}=\left(\frac{e^{\epsilon \gamma_E } s^\epsilon}{i\pi^{2-\epsilon}}\right)^2\int d^d k_1 d^d k_2 \frac{1}{D_1^{n_1}\cdots D_9^{n_9}}.\label{eq:int_def}
\end{equation}
We use \Reduze\ \cite{Studerus:2009ye,vonManteuffel:2012np} to calculate the IBP relations for this integral family and find that we need in total $37$ master integrals in order to express any other integral in terms of these. In order to find a normal basis of master intergals, we use Ref. \cite{Henn:2013woa} as a guideline to choose our pre-canonical basis. To make the comparision with that reference easier, we introduce the notation $\mathrm{f}_{n.1}$ and $\mathrm{f}_{n.2}$ , which signifies that $\mathrm{f}_{n.1}$ and $\mathrm{f}_{n.2}$ are euqal in the equal-mass-case $m_1=m_2$ and furthermore equal to the corresponding integral of Ref. \cite{Henn:2013woa}. Our pre-canonical basis is then given by
\begin{align*}
    &\mathrm{f}_{1.1}=\mathrm{I}_{0,2,0,0,0,2,0,0,0},
    &\mathrm{f}_{1.2}=\mathrm{I}_{0,0,0,0,0,2,2,0,0},\qquad
    &\mathrm{f}_{1.3}=\mathrm{I}_{2,2,0,0,0,0,0,0,0},\nonumber\\
    &\mathrm{f}_{2\phantom{.1}}=\mathrm{I}_{0,0,2,2,0,0,0,0,1},
    &\mathrm{f}_{3.1}=\mathrm{I}_{0,1,2,2,0,0,0,0,0},\qquad
    &\mathrm{f}_{3.2}=\mathrm{I}_{0,0,2,2,0,0,1,0,0}\nonumber\\
    &\mathrm{f}_{4.1}=\mathrm{I}_{1,2,0,0,0,2,0,0,0},
    &\mathrm{f}_{4.2}=\mathrm{I}_{0,2,0,0,0,2,1,0,0},\qquad
    &\mathrm{f}_{5.1}=\mathrm{I}_{0,2,2,0,0,1,0,0,0},\nonumber\\
    &\mathrm{f}_{5.2}=\mathrm{I}_{0,1,2,0,0,2,0,0,0},
    &\mathrm{f}_{6\phantom{.1}}=\mathrm{I}_{0,2,1,0,0,2,0,0,0},\qquad
    &\mathrm{f}_{7\phantom{.1}}=\mathrm{I}_{0,1,2,1,0,0,1,0,0},\nonumber\\
    &\mathrm{f}_{8.1}=\mathrm{I}_{0,2,1,1,0,1,0,0,0},
    &\mathrm{f}_{8.2}=\mathrm{I}_{1,0,1,1,0,0,2,0,0},\qquad
    &\mathrm{f}_{9.1}=\mathrm{I}_{1,0,1,1,0,0,0,0,2},\nonumber\\
    &\mathrm{f}_{9.2}=\mathrm{I}_{0,0,1,1,0,1,0,0,2},
    &\mathrm{f}_{10\phantom{.1}}=\mathrm{I}_{1,2,0,0,0,2,1,0,0},\qquad
    &\mathrm{f}_{11\phantom{.1}}=\mathrm{I}_{0,1,1,1,0,1,0,0,1},\nonumber\\
    &\mathrm{f}_{12\phantom{.1}}=\mathrm{I}_{0,1,2,1,0,1,0,0,1},
    &\mathrm{f}_{13.1}=\mathrm{I}_{0,1,1,1,0,2,0,0,1},\qquad
    &\mathrm{f}_{13.2}=\mathrm{I}_{0,2,1,1,0,1,0,0,1},\nonumber\\
    &\mathrm{f}_{14\phantom{.1}}=\mathrm{I}_{0,2,1,1,0,2,0,0,1},
    &\mathrm{f}_{15.1}=\mathrm{I}_{1,1,1,1,0,0,1,0,0},\qquad
    &\mathrm{f}_{15.2}=\mathrm{I}_{0,1,1,1,0,1,1,0,0},\nonumber\\
    &\mathrm{f}_{16.1}=\mathrm{I}_{1,1,1,1,0,0,2,0,0},
    &\mathrm{f}_{16.2}=\mathrm{I}_{0,2,1,1,0,1,1,0,0},\qquad
    &\mathrm{f}_{17.1}=\mathrm{I}_{1,1,1,1,0,0,0,0,1},\nonumber\\
    &\mathrm{f}_{17.2}=\mathrm{I}_{0,0,1,1,0,1,1,0,1},
    &\mathrm{f}_{18.1}=\mathrm{I}_{1,2,1,2,0,0,0,0,1},\qquad
    &\mathrm{f}_{18.2}=\mathrm{I}_{0,0,1,2,0,1,2,0,1},\nonumber\\
    &\mathrm{f}_{19\phantom{.1}}=\mathrm{I}_{1,0,1,1,0,1,0,0,2},
    &\mathrm{f}_{20.1}=\mathrm{I}_{1,1,1,1,0,1,0,0,1},\qquad
    &\mathrm{f}_{21.1}=\mathrm{I}_{1,1,1,1,-1,1,0,0,1},\nonumber\\
    &\mathrm{f}_{20.2}=\mathrm{I}_{1,0,1,1,0,1,1,0,1},
    &\mathrm{f}_{21.2}=\mathrm{I}_{1,0,1,1,-1,1,1,0,1},\qquad
    &\mathrm{f}_{22\phantom{.1}}=\mathrm{I}_{1,1,1,1,0,1,1,0,1},\nonumber\\
    &\mathrm{f}_{23\phantom{.1}}=\mathrm{I}_{1,1,1,1,-1,1,1,0,1}.
\end{align*}

For the convenience of the reader we also give the definition of the basis in terms of diagrams. In the following a thick line denotes a massive particle with mass $m_1$, a thin line a massive particle with mass $m_2$ and a dashed line a massless particle.
\begin{align*}
&\mathrm{f}_{1.1} = \includegraphics[valign = m, raise = 0.1 cm, scale=0.35]{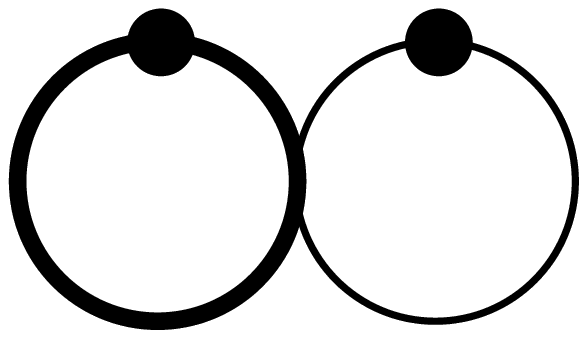}&\qquad
&\mathrm{f}_{1.2} = \includegraphics[valign = m, raise = 0.1 cm, scale=0.35]{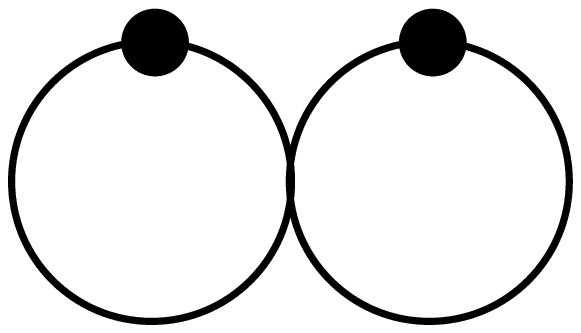}&\qquad
&\mathrm{f}_{1.3} = \includegraphics[valign = m, raise = 0.1 cm, scale=0.35]{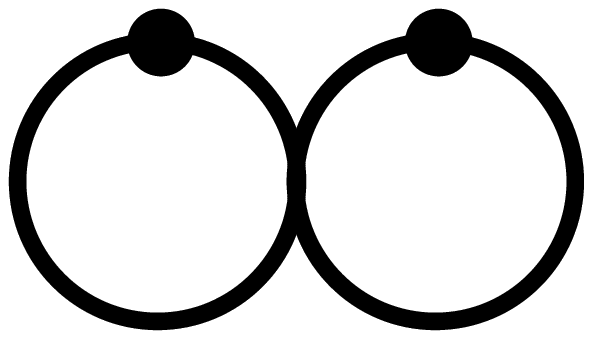} \nonumber
\\
&\mathrm{f}_{2} = \includegraphics[valign = m, raise = 0.1 cm,scale=0.35]{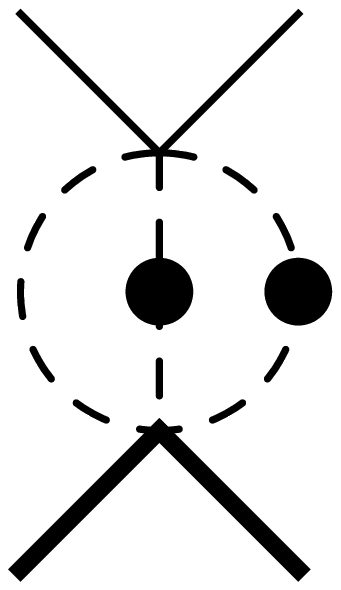}&\qquad
&\mathrm{f}_{3.1} = \includegraphics[valign = m, raise = 0.1 cm, scale=0.35]{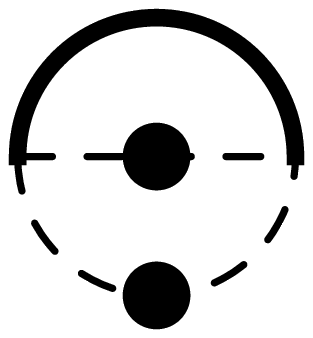}&\qquad
&\mathrm{f}_{3.2} = \includegraphics[valign = m, raise = 0.1 cm, scale=0.35]{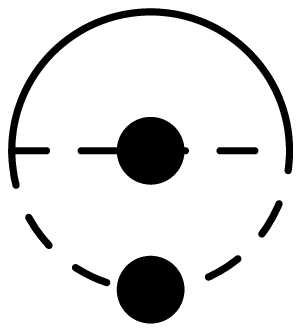} \nonumber
\\
&\mathrm{f}_{4.1} = \includegraphics[valign = m, raise = 0.1 cm, scale=0.35]{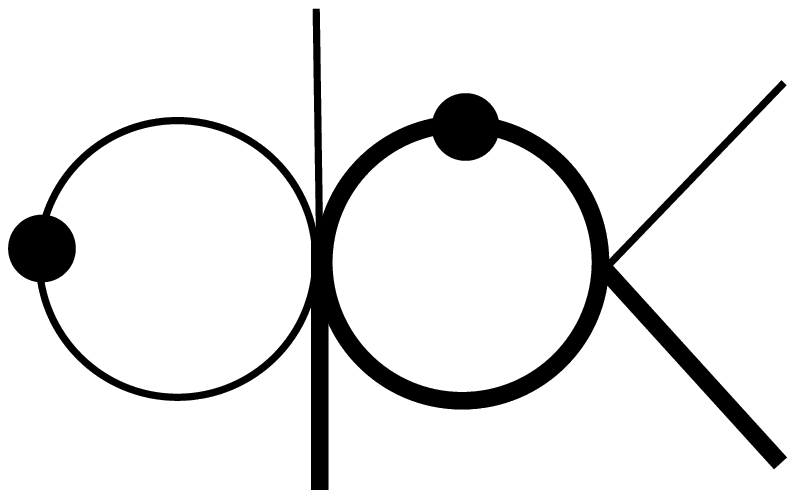}&\qquad
&\mathrm{f}_{4.2} = \includegraphics[valign = m, raise = 0.1 cm, scale=0.35]{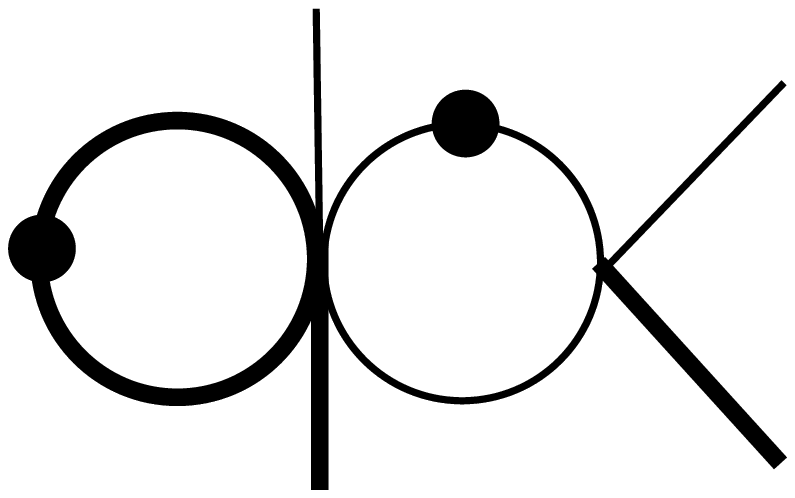}&\qquad
&\mathrm{f}_{5.1} = \includegraphics[valign = m, raise = 0.1 cm, scale=0.35]{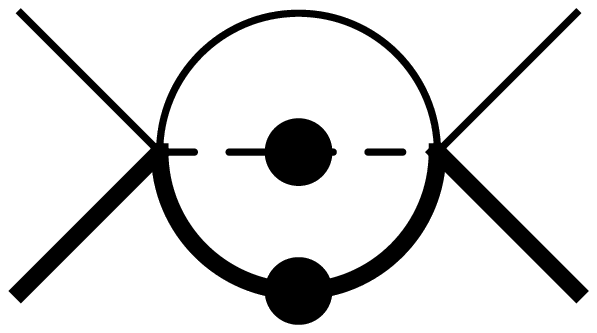} \nonumber
\\
&\mathrm{f}_{5.2} = \includegraphics[valign = m, raise = 0.1 cm, scale=0.35]{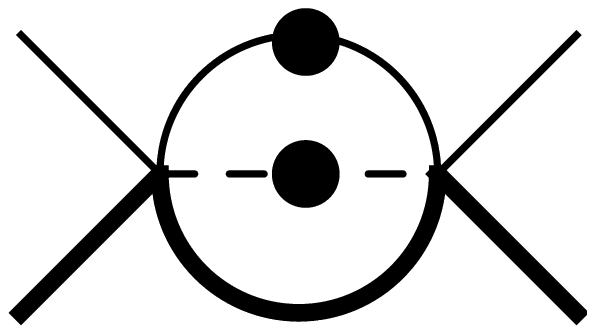}&\qquad
&\mathrm{f}_{6} = \includegraphics[valign = m, raise = 0.1 cm, scale=0.35]{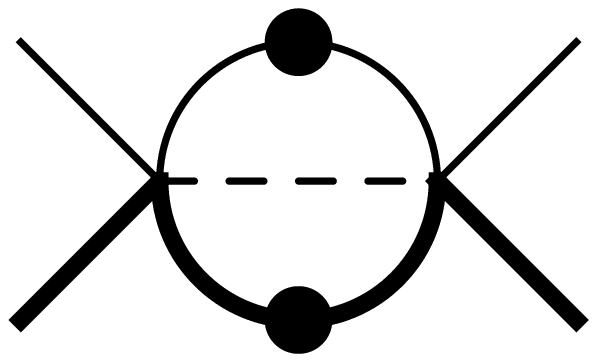}&\qquad
&\mathrm{f}_{7} = \includegraphics[valign = m, raise = 0.1 cm, scale=0.35]{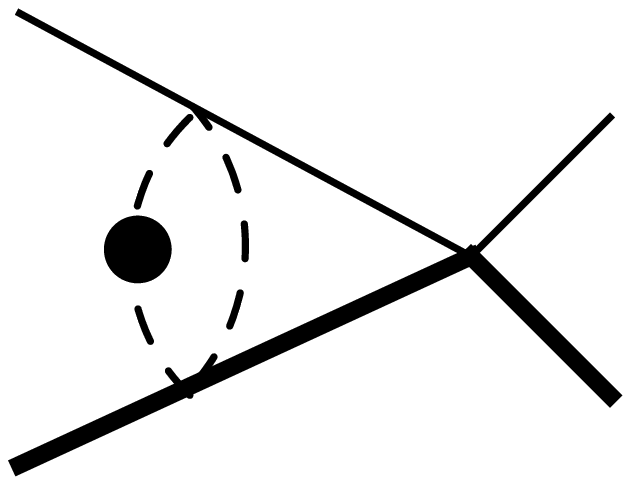} \nonumber
\\
&\mathrm{f}_{8.1} = \includegraphics[valign = m, raise = 0.1 cm, scale=0.35]{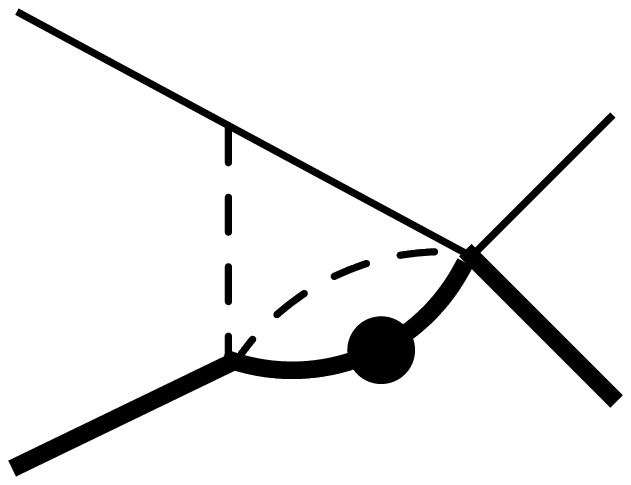}&\qquad
&\mathrm{f}_{8.2} = \includegraphics[valign = m, raise = 0.1 cm, scale=0.35]{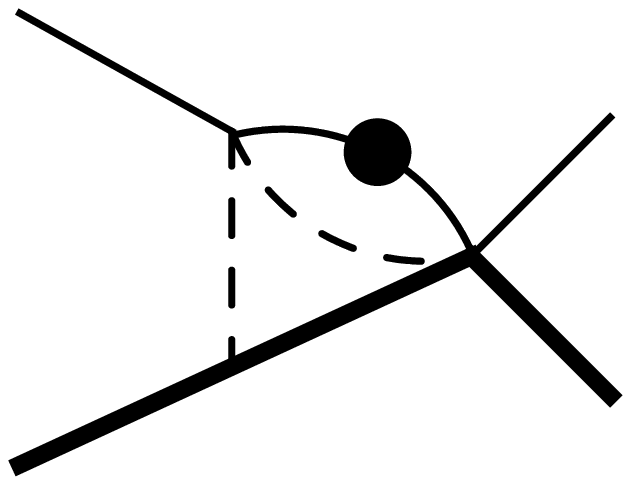}&\qquad
&\mathrm{f}_{9.1} = \includegraphics[valign = m, raise = 0.1 cm, scale=0.35, angle=180]{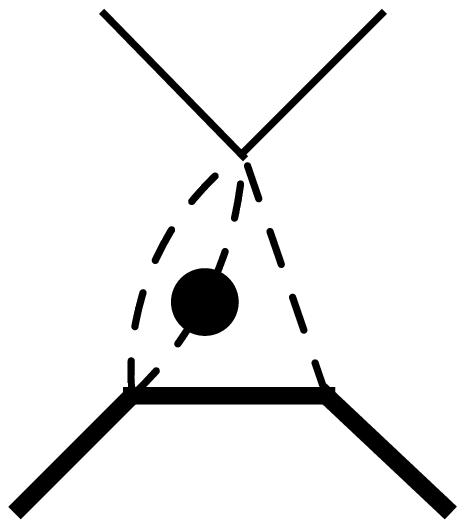}
\nonumber
\\
&\mathrm{f}_{9.2} = \includegraphics[valign = m, raise = -0.2 cm, scale=0.35]{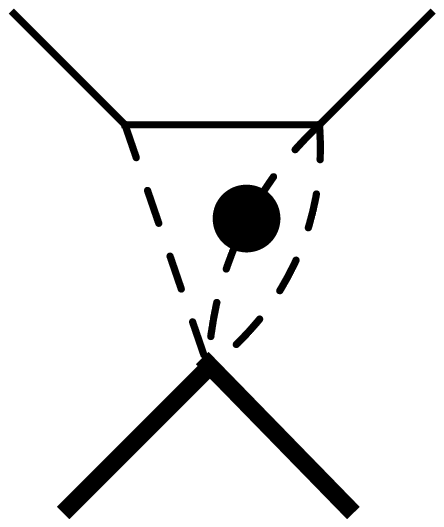}&\qquad
&\mathrm{f}_{10} = \includegraphics[valign = m, raise = 0.1 cm, scale=0.35]{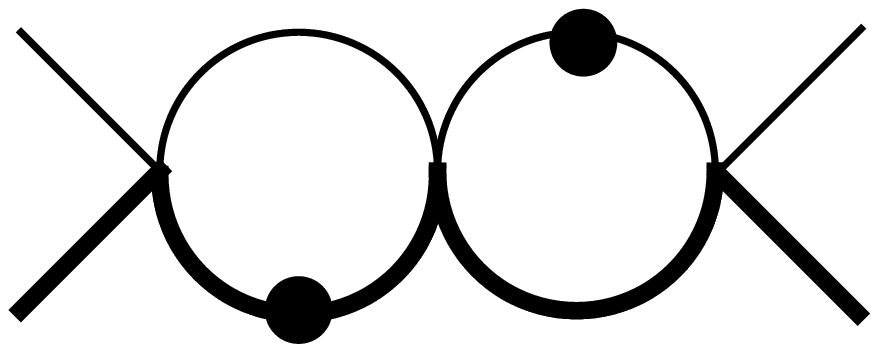}&\qquad
&\mathrm{f}_{11} = \includegraphics[valign = m, raise = 0.1 cm, scale=0.35]{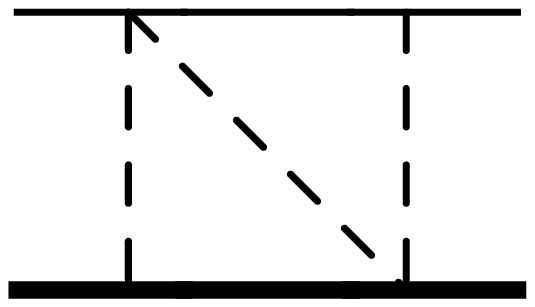}
\nonumber
\\
&\mathrm{f}_{12} = \includegraphics[valign = m, raise = 0.1 cm, scale=0.35]{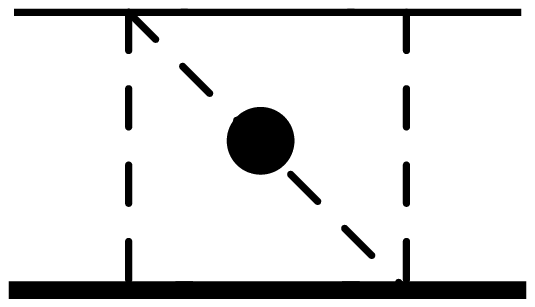}&\qquad
&\mathrm{f}_{13.1} = \includegraphics[valign = m, raise = 0.1 cm, scale=0.35]{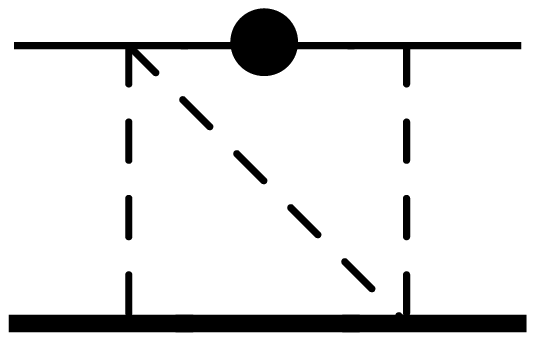}&\qquad
&\mathrm{f}_{13.2} = \includegraphics[valign = m, raise = 0.1 cm, scale=0.35]{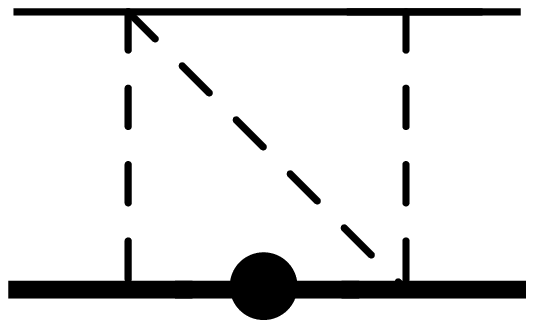}
\nonumber
\\
&\mathrm{f}_{14} = \includegraphics[valign = m, raise = 0.1 cm, scale=0.35]{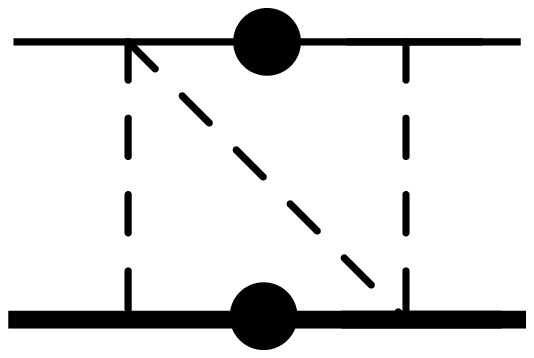}&\qquad
&\mathrm{f}_{15.1} = \includegraphics[valign = m, raise = 0.1 cm, scale=0.35]{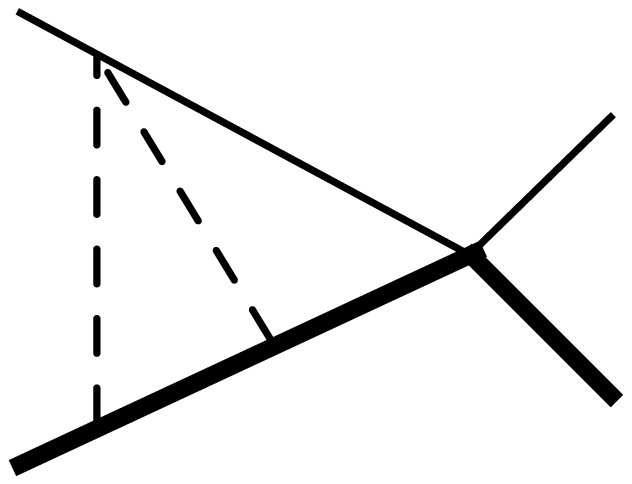}&\qquad
&\mathrm{f}_{15.2} = \includegraphics[valign = m, raise = 0.1 cm, scale=0.35]{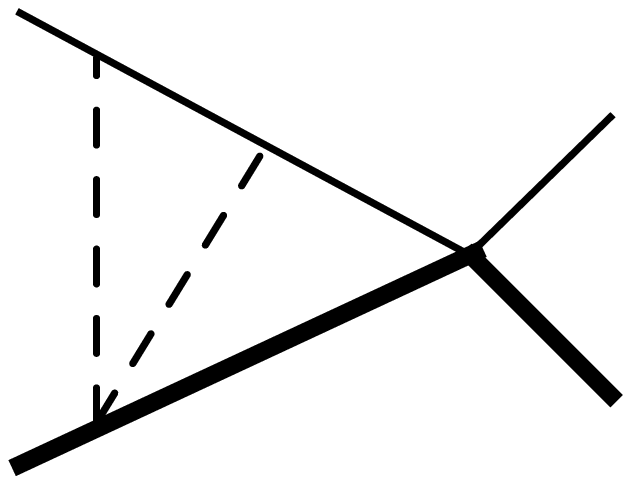}
\nonumber
\\
&\mathrm{f}_{16.1} = \includegraphics[valign = m, raise = 0.1 cm, scale=0.35]{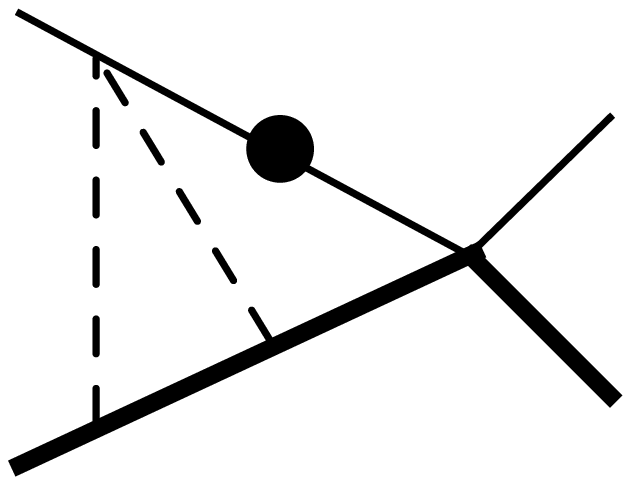}&\qquad
&\mathrm{f}_{16.2} = \includegraphics[valign = m, raise = 0.1 cm, scale=0.35]{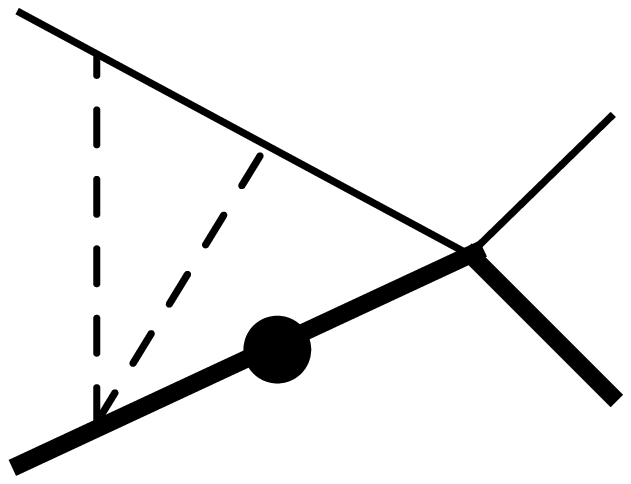}&\qquad
&\mathrm{f}_{17.1} = \includegraphics[valign = m, raise = 0.1 cm, scale=0.35, angle=180]{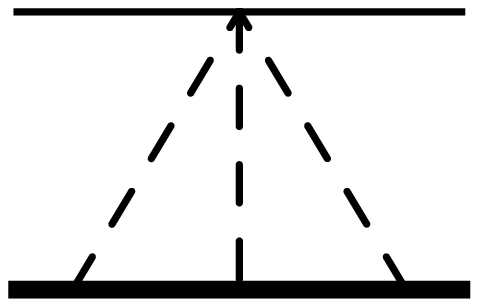}
\nonumber
\\
&\mathrm{f}_{17.2} = \includegraphics[valign = m, raise = -0.2 cm, scale=0.35]{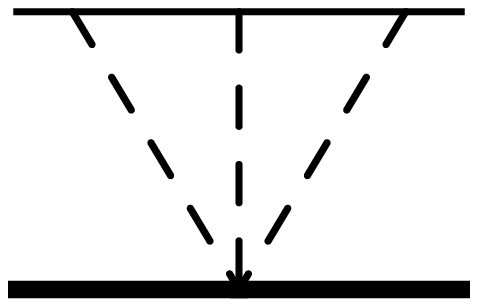}&\qquad
&\mathrm{f}_{18.1} = \includegraphics[valign = m, raise = 0.2 cm, scale=0.35, angle=180]{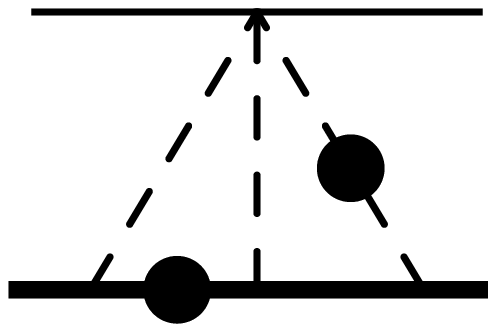}&\qquad
&\mathrm{f}_{18.2} = \includegraphics[valign = m, raise = -0.2 cm, scale=0.35]{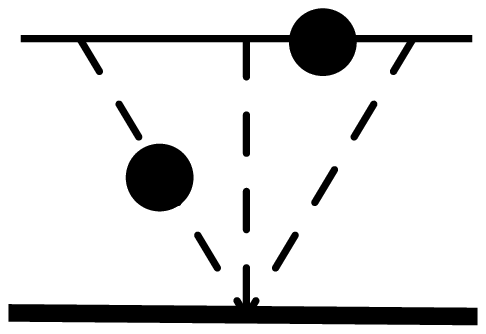}
\nonumber
\\
&\mathrm{f}_{19\phantom{.1}} = \includegraphics[valign = m, raise = 0.1 cm, scale=0.35]{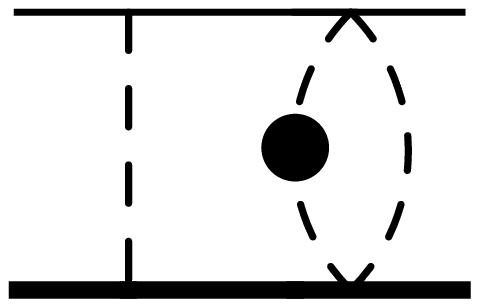}&\qquad
&\mathrm{f}_{20.1} = \includegraphics[valign = m, raise = 0.1 cm, scale=0.35]{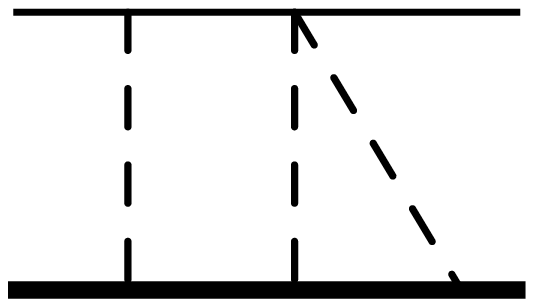}&\qquad
&\mathrm{f}_{20.2} = \includegraphics[valign = m, raise = 0.1 cm, scale=0.35]{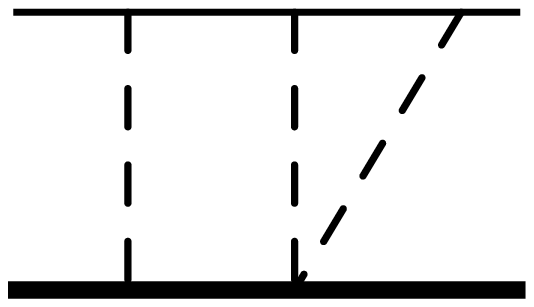} \nonumber
\\
&\mathrm{f}_{21.1} = \includegraphics[valign = m, raise = 0.1 cm, scale=0.35]{figures_bhabha/int_20_1.eps}[(k_2-p_1)^2]&\qquad
&\mathrm{f}_{21.2} = \includegraphics[valign = m, raise = 0.1 cm, scale=0.35]{figures_bhabha/int_20_2.eps}[(k_2-p_1)^2]&\qquad
&\mathrm{f}_{22\phantom{.1}} = \includegraphics[valign = m, raise = 0.1 cm, scale=0.35]{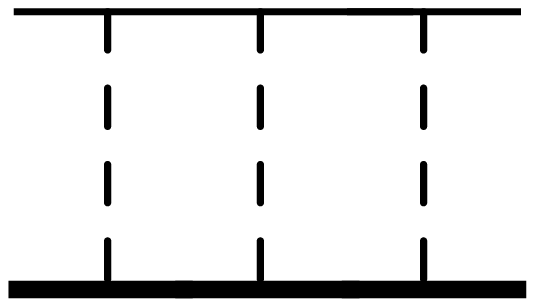}\nonumber\\
&\mathrm{f}_{23\phantom{.1}} = \includegraphics[valign = m, raise = 0.1 cm, scale=0.35]{figures_bhabha/int23} [(k_2-p_1)^2]
\end{align*}

We are now able to define the canonical basis. Again, we use Ref \cite{Henn:2013woa} as a guideline and use the fact that our basis coincides with their basis (up to factors of $1/2$ or $1/4$ for some integrals) in the equal-mass case. 
\begin{align}
    \mathrm{m}_{1.1}&=\epsilon^2\;\mathrm{f}_{1.1},\nonumber\\
    \mathrm{m}_{1.2}&=\epsilon^2\;\mathrm{f}_{1.2},\nonumber\\
    \mathrm{m}_{1.3}&=\epsilon^2\;\mathrm{f}_{1.3},\nonumber\\
    \mathrm{m}_{2}&=\epsilon^2 t\;\mathrm{f}_{2},\nonumber\\
    \mathrm{m}_{3.1}&=\frac{\epsilon^2(1+4\epsilon)m_1^2}{(1+\epsilon)}\;\mathrm{f}_{3.1},\nonumber\\
    \mathrm{m}_{3.2}&=\frac{\epsilon^2(1+4\epsilon)m_2^2}{(1+\epsilon)}\;\mathrm{f}_{3.2},\nonumber\\
    \mathrm{m}_{4.1}&=\frac{\epsilon^2\sqrt{m_1^4+(m_2^2-s)^2-2m_1^2(m_2^2+s)}\;s}{(m_1^2-m_2^2+s)}\left(\mathrm{f}_{3.1}+\frac{1}{2s\epsilon^2}\mathrm{m}_{1.3}-\frac{1}{2s\epsilon^2}\mathrm{m}_{1.1}\right),\nonumber\\
    \mathrm{m}_{4.2}&=\frac{\epsilon^2\sqrt{m_1^4+(m_2^2-s)^2-2m_1^2(m_2^2+s)}\;s}{(m_2^2-m_1^2+s)}\left(\mathrm{f}_{3.2}+\frac{1}{2s\epsilon^2}\mathrm{m}_{1.2}-\frac{1}{2s\epsilon^2}\mathrm{m}_{1.1}\right),\nonumber\\
    \mathrm{m}_{6}&=\epsilon^2 s\;\mathrm{f}_6,\nonumber\\
    \mathrm{m}_{7}&=-2\epsilon^3\sqrt{m_1^4+(m_2^2-s)^2-2m_1^2(m_2^2+s)}\;\mathrm{f}_7,\nonumber\\
    \mathrm{m}_{8.1}&=2\epsilon^3\sqrt{m_1^4+(m_2^2-s)^2-2m_1^2(m_2^2+s)}\;\mathrm{f}_{8.1},\nonumber\\
    \mathrm{m}_{8.2}&=2\epsilon^3\sqrt{m_1^4+(m_2^2-s)^2-2m_1^2(m_2^2+s)}\;\mathrm{f}_{8.2},\nonumber\\
    \mathrm{m}_{9.1}&=-2\epsilon^3\sqrt{-t(4m_1^2-t)}\;\mathrm{f}_{9.1},\nonumber\\
    \mathrm{m}_{9.2}&=-2\epsilon^3\sqrt{-t(4m_2^2-t)}\;\mathrm{f}_{9.2},\nonumber\\
    \mathrm{m}_{10}&=\frac{\epsilon^2s^2(s-(m_1+m_2)^2)(s-(m_1-m_2)^2)}{(m_1^2-m_2^2+s)(-m_1^2+m_2^2+s)}\mathrm{f}_{10}+\left(\frac{m_1^2}{m_1^2-m_2^2+s}+\frac{m_2^2}{m_2^2-m_1^2+s}\right)\mathrm{m}_{1.1}\nonumber\\
    &\phantom{=}-\left(\frac{m_1^2}{2(m_1^2-m_2^2+s)}+\frac{m_2^2}{2(m_2^2-m_1^2+s)}\right)\mathrm{m}_{1.2}-\left(\frac{m_1^2}{2(m_1^2-m_2^2+s)}+\frac{m_2^2}{2(m_2^2-m_1^2+s)}\right)\mathrm{m}_{1.3}\nonumber\\
    &\phantom{=}+\frac{(m_1^2-m_2^2)\sqrt{m_1^4+(m_2^2-s)^2-2m_1^2(m_2^2+s)}}{(m_1^2-m_2^2)^2-s^2}\;\mathrm{m}_{4.1}\nonumber\\
    &\phantom{=}+\frac{(m_2^2-m_1^2)\sqrt{m_1^4+(m_2^2-s)^2-2m_1^2(m_2^2+s)}}{(m_1^2-m_2^2)^2-s^2}\;\mathrm{m}_{4.2}\nonumber\\
    \mathrm{m}_{11}&=4\epsilon^4\sqrt{m_1^4+(-m_2^2+s+t)^2-2m_1^2(m_2^2+s+t)}\;\mathrm{f}_{11},\nonumber\\
    \mathrm{m}_{12}&=-2\epsilon^3\sqrt{m_1^4+(-m_2^2+s+t)^2-2m_1^2(m_2^2+s+t)}\;t\; \mathrm{f}_{12},\nonumber\\
    \mathrm{m}_{13.1}&=-2\epsilon^3 m_2^2 \sqrt{-t(4m_1^2-t)}\;\mathrm{f}_{13.1},\nonumber\\
    \mathrm{m}_{13.2}&=-2\epsilon^3 m_1^2 \sqrt{-t(4m_2^2-t)}\;\mathrm{f}_{13.2},\nonumber\\
    \mathrm{m}_{15.1}&=4\epsilon^4\sqrt{m_1^4+(-m_2^2+s+t)^2-2m_1^2(m_2^2+s+t)}\;\mathrm{f}_{15.1},\nonumber\\
    \mathrm{m}_{15.2}&=4\epsilon^4\sqrt{m_1^4+(-m_2^2+s+t)^2-2m_1^2(m_2^2+s+t)}\;\mathrm{f}_{15.2},\nonumber\\
    \mathrm{m}_{16.1}&=\epsilon^3(m_1^2-2m_1 m_2+m_2^2-s)(m_1^2+2m_1 m_2+m_2^2-s)\;\mathrm{f}_{16.1},\nonumber\\
    \mathrm{m}_{16.2}&=\epsilon^3(m_1^2-2m_1 m_2+m_2^2-s)(m_1^2+2m_1 m_2+m_2^2-s)\;\mathrm{f}_{16.2},\nonumber\\
    \mathrm{m}_{17.1}&=4\epsilon^4\sqrt{-t(4m_1^2-t)}\mathrm{f}_{17.1},\nonumber\\
    \mathrm{m}_{17.2}&=4\epsilon^4\sqrt{-t(4m_2^2-t)}\mathrm{f}_{17.2},\nonumber\\
    \mathrm{m}_{18.1}&=\epsilon(1+2\epsilon)t m_1^4 \mathrm{f}_{18.1}+\frac{t}{2\sqrt{-t(4m_1^2-t)}}\mathrm{m}_{17.1}+\frac{t-3m_1^2}{\sqrt{-t(4m_1^2-t)}}\mathrm{m}_{9.1}+\frac{1+\epsilon}{1+4\epsilon}\mathrm{m}_{3.1},\nonumber\\
    \mathrm{m}_{18.2}&=\epsilon(1+2\epsilon)t m_2^4 \mathrm{f}_{18.2}+\frac{t}{2\sqrt{-t(4m_2^2-t)}}\mathrm{m}_{17.2}+\frac{t-3m_2^2}{\sqrt{-t(4m_1^2-t)}}\mathrm{m}_{9.2}+\frac{1+\epsilon}{1+4\epsilon}\mathrm{m}_{3.2},\nonumber\\
    \mathrm{m}_{19}&=\epsilon^3\sqrt{m_1^4+(m_2^2-s)^2-2m_1^2(m_2^2+s)}\;t\;\mathrm{f}_{19},\nonumber\\
    \mathrm{m}_{20.1}&=\epsilon^4\sqrt{m_1^4+(m_2^2-s)^2-2m_1^2(m_2^2+s)}\sqrt{-t(4m_1^2-t)}\;\mathrm{f}_{20.1},\nonumber\\
    \mathrm{m}_{20.2}&=\epsilon^4\sqrt{m_1^4+(m_2^2-s)^2-2m_1^2(m_2^2+s)}\sqrt{-t(4m_2^2-t)}\;\mathrm{f}_{20.2},\nonumber\\
    \mathrm{m}_{21.1}&=\epsilon^3(2\epsilon-1)\sqrt{m_1^4+(m_2^2-s)^2-2m_1^2(m_2^2+s)}\mathrm{f}_{21.1}\nonumber\\
    &\phantom{=}-\frac{2t}{\sqrt{-t(4m_1^2-t)}}\mathrm{m}_{20.1}-\frac{\sqrt{m_1^4+(m_2^2-s)^2-2m_1^2(m_2^2+s)}}{2\sqrt{m_1^4+(-m_2^2+s+t)^2-2m_1^2(m_2^2+s+t)}}\mathrm{m}_{11},\nonumber\\
    \mathrm{m}_{21.2}&=\epsilon^3(2\epsilon-1)\sqrt{m_1^4+(m_2^2-s)^2-2m_1^2(m_2^2+s)}\mathrm{f}_{21.2}\nonumber\\
    &\phantom{=}-\frac{2t}{\sqrt{-t(4m_2^2-t)}}\mathrm{m}_{20.2}-\frac{\sqrt{m_1^4+(m_2^2-s)^2-2m_1^2(m_2^2+s)}}{2\sqrt{m_1^4+(-m_2^2+s+t)^2-2m_1^2(m_2^2+s+t)}}\mathrm{m}_{11},\nonumber\\
    \mathrm{m}_{22}&=\epsilon^4 t (m_1^4+(m_2^2-s)^2-2m_1^2(m_2^2+s))\;\mathrm{f}_{22},\nonumber\\
    \mathrm{m}_{23}&=\epsilon^4 (m_1^4+(m_2^2-s)^2-2m_1^2(m_2^2+s))\;\mathrm{f}_{23}.\label{eq:masters}
\end{align}

The integrals $\vec{\mathrm{m}}$ fullfill a canonical differential equation, i.e. a differential in which the dimensional regulator $\epsilon$ decouples from the kinematical part, and which is in $\ud\ln$ form, i.e.
\begin{equation}
    \vec{\mathrm{m}}=\epsilon \sum_{i}A_i\ud\ln(\tilde l_i)  \vec{\mathrm{m}}.
\end{equation}
Here, $A_i$ are matrices with rational numbers as entries and $\tilde l_i$ are algebraic functions depending on $x$, $y$ and $z$. The differential equation is to lengthy to write it down explicitly, instead we give it in electronic form in an ancillary file.

Note that in total $4$ square roots appear in the definition of the normal form master integrals, and therefore also in the differential equation. In order to proceed and integrate the differential equation, one can try two different approaches. In the first one, one tries to find a reparametrization which rationalizes (some of) the square roots \cite{Besier:2018jen}. In that way one can often directly integrate the differential equation in terms of GPL`s, if one encounters linear letter. In the present case, it is proven that at least one square root is unrationalizable, such that a direct integration approach in terms of GPL`s does not work \cite{Besier:2019hqd}. In the second approach, one tries to match the symbol of the differential equation against a suitable ansatz of multiple polylogarithm \cite{Heller:2019gkq}. In this approach one can of course also try to rationalize as many roots as possible before the matching as was done in Ref \cite{Heller:2019gkq}. However, in principle, this is not necessary anymore.

In this work, we will do the second approach, without rationalizing any root, i.e. we deal with all four at the same time when matching the differential equation against an ansatz.

\section{Simplifying the differential equation}
\label{sec:deq_simple}

As mentioned in the previous section, with the master integrals defined in Eq. \eqref{eq:masters}, we are able to derive a canonical differential equation in $\epsilon$-decoupled, $\ud\ln$ form \cite{Henn:2013pwa,Kotikov:2010gf}:
\begin{equation}
    \vec{\mathrm{m}}=\epsilon \sum_{i=1}^{30}A_i\ud\ln(\tilde l_i)  \vec{\mathrm{m}}.
\end{equation}
At first sight, we find in this differential equation in total $30$ letter, which we list here:
\begin{align}
    \tilde{\mathcal{L}}=\bigg\{&x-y,x+y,y,2 y+1,2 y+z+1,2 y-1,2 y-z-1,z,2 y+\sqrt{4 y^2-4 x+1}-1,\nonumber\\
    &2 y+\sqrt{4 y^2-4 x+1}+1,\sqrt{4 y^2-4 x+1},\sqrt{4 x+4 y-z}+\sqrt{-z},\sqrt{4 x+4 y-z},\nonumber\\
   &\sqrt{4 x-4 y-z}+\sqrt{-z},\sqrt{4
   x-4 y-z},2 y-z+\sqrt{4 y^2+z^2-4 x-4 x z+2 z+1}-1,\nonumber\\
   &2 y+z+\sqrt{4 y^2+z^2-4 x-4 x z+2 z+1}+1,-4 y^2+4 x-z-1,4 y^2-z-1,\nonumber\\
   &-4 y^2+2 z y+4
   x-z+\sqrt{4 y^2-4 x+1} \sqrt{4 y^2+z^2-4 x-4 x z+2 z+1}-1,\nonumber\\
   &-4 y^2-2 z y+4 x-z+\sqrt{4 y^2-4 x+1} \sqrt{4 y^2+z^2-4 x-4 x z+2 z+1}-1,\nonumber\\
   &-4 \sqrt{-z}
   x+\sqrt{-z} z-2 y \sqrt{-z}+\sqrt{-z}+\sqrt{4 x+4 y-z} \sqrt{4 y^2+z^2-4 x-4 x z+2 z+1},\nonumber\\
   &-4 \sqrt{-z} x+\sqrt{-z} z+2 y \sqrt{-z}+\sqrt{-z}+\sqrt{4 x-4 y-z} \sqrt{4
   y^2+z^2-4 x-4 x z+2 z+1},\nonumber\\
   &2 \sqrt{-z} y+\sqrt{4 y^2-4 x+1} \sqrt{4 x+4 y-z}+\sqrt{-z},4 y^2+4 y-z+1,\nonumber\\
   &-8 y^3+8 x y^2-4 z y^2+8 y^2-8 x
   y+2 z y-2 \sqrt{4 y^2-4 x+1} \sqrt{z} \sqrt{-4 x+4 y+z} y\nonumber\\
   &-2 y+2 x+2 x z-z+\sqrt{4 y^2-4 x+1} \sqrt{z} \sqrt{-4 x+4 y+z},\nonumber\\
   &-16 y^4+16 y^3+16 x y^2-8 y^2-16 x y-4 z
   y+4 \sqrt{4 y^2-4 x+1} \sqrt{z} \sqrt{-4 x+4 y+z} y\nonumber\\
   &+4 y-z^2+4 x+4 x z-2 \sqrt{4 y^2-4 x+1} \sqrt{z} \sqrt{-4 x+4 y+z}-1,\nonumber\\
   &4 x-z+\sqrt{-4 x-4 y+z} \sqrt{-4 x+4
   y+z}-2,\sqrt{4 y^2+z^2-4 x-4 x z+2 z+1},\nonumber\\
   &2 \sqrt{-z} y+\sqrt{4 y^2-4 x+1} \sqrt{4 x-4 y-z}-\sqrt{-z}\bigg\}.\label{eq:letter_first}
\end{align}
Clearly, the structure of the letter is very complicated. To further process the differential equation and match it against a suitable ansatz, we need to simplify the letter. We weill du this along the same lines as discussed in Ref. \cite{Heller:2019gkq}. First, we identify the rational part of the alphabet and note that there are $4$ roots, which are given by:
\begin{align}
    r_1&=\sqrt{1 - 4 x + 4 y^2},\quad r_2=\sqrt{-4 x z + 4 y z + z^2},\quad r_3=\sqrt{-4 x z - 4 y z + z^2},\nonumber\\
    r_4&=\sqrt{1 - 4 x + 4 y^2 + 2 z - 4 x z + z^2}.
\end{align}
Note that we do not use $\sqrt{-z}$ as separate root, but rather prefer to not factorize $r_2$ and $r_3$. This choice seems natural to us, since $\sqrt{-t}$ does not appear as separate root in the definition of the master integrals in Eq. \eqref{eq:masters}. 

Our starting point for the construction of new and simpler letter is now to consider the rational part of the alphabet, given by
\begin{align}
    \tilde{\mathcal{L}}_R=\bigg\{&x-y,x+y,y,2 y-1,2 y+1,z+1,4 x-4 y^2-z-1,z,4 y^2-z-1,2 y+z+1,\nonumber\\
    &2 y-z-1,4  y^2+4 y-z+1,4 y^2-4 y-z+1\bigg\}.\label{eq:letter_rat}
\end{align}
As explained in Ref. \cite{Heller:2019gkq}, we proceed by making an ansatz to find simple algebraic letter, which factorize over the rational alphabet, i.e. for all $4$ roots we try to find letter of the form
\begin{equation}
    l=q(x,y,z)+r_i,\qquad \bar{l}=q(x,y,z)-r_i,
\end{equation}
such that $\bar{l}l$ factorizes over the functions in Eq. \eqref{eq:letter_rat}, and where $q$ is a polynomial in $x$, $y$ and $z$. In this way, we find the following candidates:
\begin{align*}
\bigg\{&\frac{r_1}{2}+\frac{1}{2}(-1+2x),\frac{r_1}{2}+\frac{1}{2}(-1+2y),\frac{r_1}{2}+\frac{1}{2}(1+2y),\frac{r_2}{2}+\frac{1}{2}(2x-2y-z),\nonumber\\
&\frac{r_2}{2}+\frac{z}{2},\frac{r_3}{2}+\frac{1}{2}(2x+2y-z),\frac{r_3}{2}+\frac{z}{2},\frac{1}{2}(-1 + 2 x - z) + \frac{r_4}{2},\dots\bigg\}.
\end{align*}
It turns out that using only these candidates we do not find enough new, easier letter to re-express all old letter from Eq. \eqref{eq:letter_first} in terms of them. Therefore, we need to include more terms in our search. To do this, we proceed by making another ansatz with two roots,
\begin{equation}
     l=q(x,y,z)+r_ir_j,\qquad \bar{l}=q(x,y,z)-r_ir_j,
\end{equation}
such that again $\bar{l}l$ factorizes over the functions in Eq. \eqref{eq:letter_rat}. This generalization to simplify letters beyond one root was already used in appendix B in Ref. \cite{Heller:2019gkq}. 

In the present calculation, we discover that some of the candidates we find with this extended ansatz can be reduced even further. For example, we find that
\begin{equation}
    l=\frac{1}{2}\left(-z + 2 y z+r_1 r_2\right)
\end{equation}
is a valid candidate, since 
\begin{equation}
    \frac{1}{2}\left(-z + 2 y z+r_1 r_2\right)\frac{1}{2}\left(-z + 2 y z-r_1 r_2\right)=-(x - y) (-1 + 4 x - 4 y^2 - z) z.
\end{equation}
However, it is possible to factorize that candidate further taking into account another root, $r_4$. In particular, we have
\begin{equation}
    \frac{1}{2}\left(-z + 2 y z+r_1 r_2\right)=\frac{1}{2}(r_1 + r_2 + r_4) \frac{1}{2}(r_1 + r_2 - r_4),
\end{equation}
which gives us two better and easier candidates, i.e. $(r_1 + r_2 + r_4)/2$ and $(r_1 + r_2 - r_4)/2$. We therefore always try to simplify a given candidate with two roots further along the same lines as above as a sum of three roots.

Using these ideas, we are able to simplify the initial alphabet drastically. Our final alphabet is given by the rational part
\begin{equation}
   \mathcal{L}_R=\bigg\{x - y, x + y, -1 + 4 x - 4 y^2 - z, z\bigg\}\label{eq:letter_rat_easy}
\end{equation}
and the algebraic part
\begin{align}
 \mathcal{L}_A=\bigg\{&r_1, 
     r_2, 
     r_3, 
     r_4, 
 \frac{1}{2} \big(-1 + 2 y + r_1\big), 
 \frac{1}{2} \big(1 + 2 y + r_1\big), 
 \frac{1}{2} \big(r_2 + z\big), 
 \frac{1}{2} \big(r_3 + z\big), \nonumber\\
&\frac{1}{2} \big(-1 + 2 x - z + r_4\big),
 \frac{1}{2} \big(r_2 + r_1 + r_4\big), 
 \frac{1}{2} \big(r_3 + r_1 + r_4\big), 
 \frac{1}{2} \big(r_2 - r_1 - r_4\big),\nonumber\\
&\frac{1}{2} \big(r_3 - r_1 - r_4\big),
 \frac{1}{2} \big(r_1 - r_3 - r_4\big),
 \frac{1}{2} \big(r_1 - r_2 - r_4\big), 
 \frac{1}{2} \big(z (-2 + 4 x - z) + r_2 r_3\big)\bigg\}.\label{eq:letter_alg_easy}
\end{align}
Note, that the new alphabet has a remarkable simple structure. Especially the structure of the letter which are sums of three roots seems much more natural compared to the initial alphabet in Eq. \eqref{eq:letter_first}. Note also, that the number of letter was reduced from $30$ in the starting alphabet to $20$ in the final one. This shows that it is quite non-trivial to see if a given differential equation in $\ud\ln$ form has a minimal number of letter or not.

\section{Integrating the symbol and matching}
\label{sec:integration}

In order to integrate the differential equation, we follow the lines of Ref. \cite{Duhr:2011zq}. The idea behind this approach is to construct candidate functions whose symbols can be matched against the symbol of the differential equation at a given weight. At weight one only logarithms are needed. For higher weights, one uses ${\rm Li}$ functions, whose symbol is defined
\begin{equation}
\mathcal{S}\big({\rm Li}_n(f)\big)=-(1-f)\otimes \underbrace{f\otimes...\otimes f}_{(n-1)~\text{times}}.\label{eq:symbol}
\end{equation}
From Eq. \eqref{eq:symbol} it is clear that in order to not introduce new, spurious letter, we need to make sure that both $f$ and $1-f$ factorize over the symbol alphabet of the differential equation. In practice, one constructs power-products $f$ of the letter and checks that $1-f$ factorizes over the alphabet. In addition to  ${\rm Li}_n$ functions, one also needs to consider functions of higher depth. To treat ${\rm Li}_{2,1}$, ${\rm Li}_{3,1}$, and ${\rm Li}_{2,2}$, one can again check their symbols. In order to not introduce new, spurious letter, the pair of function arguments $(f_i,f_j)$ of a ${\rm Li}_{n_1,n_2}$ function has to be chosen such that $1-f_i f_j$ factorizes over the alphabet, and where $f_i$ and $f_j$ can either be $1$, or chosen from the set of admissible ${\rm Li}_n$ function arguments.

In order to check if for a given power product $f$ of letter also $1-f$ factorizes over the alphabet, we use the same approach as in Ref. \cite{Heller:2019gkq} - for a given expression $g$ we are interested in factorizations of the form
\begin{equation}
\label{powerproduct}
g = c^{a_0} l_1^{a_1} l_2^{a_2} \cdots,
\end{equation}
with a rational number $c$. Due to the presence of the square roots in the alphabet, in general one needs to consider half-integer exponents (roots) or, even worse, rational number exponents of letters in Eq. \eqref{powerproduct}, such that $a_n\in \mathbbm{Q}$. However, we observe that using a simplified alphabet, as given by Eqs. \eqref{eq:letter_rat_easy} and \eqref{eq:letter_alg_easy} it is sufficient to use integer exponents, as in the rational case.

It is a non-trivial task to find factorizations of the form as in Eq. \eqref{powerproduct} using standard computer algebra systems due to the presence of the four roots $r_1,\dots r_4$ in the letter. To deal with that issue we use a heuristic factorization approach: we observe, that the factorization \eqref{powerproduct} implies
\begin{equation}
\label{logrelations}
\ln(g) - a_0 \ln(c) - a_1 \ln(l_1) - a_2 \ln(l_2) - \ldots = 0\,.
\end{equation}
Replacing the variables by numerical samples allows us to find such relations using integer relation finders.
To find the required factorizations we use the {\tt C++} program \texttt{Hactor} \cite{hactor}, which is based on the Lenstra-Lenstra-{Lov\'asz} (LLL)
algorithm~\cite{Lenstra1982} implemented in {\tt PARI/GP}~\cite{PARI2}.

In the present case we find in total $1057$ admissible $\mathrm{Li}_n$ and $29015$ admissible $\mathrm{Li}_{n,m}$ function arguments. In order to make the ansatz smaller, we restrict ourselves only to functions which are real-valued in the physical phase-space. For example, for a given argument $f$ for of a  $\mathrm{Li}_n$ function, this means that we demand 
\begin{equation}
    f<1\quad \mathrm{for}\quad 0<y<\frac{1}{2},\quad y<x<\frac{1}{4}\big(1+4y^2\big),\quad -1+4x-4y^2<z<0.\label{eq:check_li}
\end{equation}
For most of the function arguments we were able to check Eq. \eqref{eq:check_li} using \texttt{Reduce} in \texttt{Mathematica}. However, some of the candidates are to complex to check this property analytically. In these cases, we used numerical samples to check Eq. \eqref{eq:check_li}. With the resulting set of $\mathrm{Li}$ functions we were able to match the symbol of the differential equations against a linear combination of the set using the \texttt{Mathematica} package \texttt{GPL.m} \cite{GPL}, which uses \texttt{GiNaC} \cite{Bauer:2000cp,Vollinga:2004sn} for the numerical evaluation of multiple polylogarithms.

The next step is to determine the boundary constants which result from the integration. Since all integrals were known for the equal-mass case from Ref. \cite{Henn:2013woa}, we matched our solution against the solution of that reference at a phase-space point in the physical region. To do this, we evaluated both solutions using \texttt{GiNaC} with a precision of up to $150$ digits. Using the PSLQ algorithm \cite{PSLQ}, we were able to fit all boundary constants against an ansatz made from the following set of constants:
\begin{equation}
    \pi,\;\ln(2),\;\zeta_3,\;\mathrm{Li_2\left(\frac{1}{2}\right)}.
\end{equation}
Note that $\mathrm{m}_{11}$ is the only integral which was not given in analytic form in Ref. \cite{Henn:2013woa}, since it is the only integral in the equal-mass case in which a unrationalizable square root appears at weight 4. We therefore matched the boundary constants of $\mathrm{m}_{11}$ such that
\begin{equation}
    \mathrm{m}_{11}(x=1/4,y=0,z=0)=0,
\end{equation}
which corresponds to the two-particle threshold for equal masses. Note that the prefactor of this integral is zero at that point.

After the matching, we find in total $2211$ different functions in the final result. Let us stress again that all functions that appear in the solution are chosen such that they are real-valued in the physical phase-space and have no pseudo-thresholds in the whole region.

\section{Results and numerical evaluation}
\label{sec:checks}
We append the analytic solution to all $37$ master integrals in electronic form as ancillary files in the \texttt{arXiv} submission. We provide in total four files: \texttt{ints.m} contains the anlytic solution of all $37$ master integrals, \texttt{intdefs.m} contains the defintion of the canonical basis as given in Eq. \eqref{eq:masters} and \texttt{dlog.m} and \texttt{dlog-simple.m} both contain the differential equation in d-log form, once in terms of the complicated alphabet containing $30$ letter and once in the simple form containing only $20$ letter.

We give explicit numeric results at the phase-space point $s=30$, $t=-5$, $m_1=2$, and $m_2=1$, which corresponds to the point
\begin{equation}
    x=\frac{1}{12},\quad y=\frac{1}{20},\quad z=-\frac{1}{6},
\end{equation}
for the last two integrals.
\begin{align*}
   \mathrm{m}_{22}\;\approx& \;(-7.045803264961093967\dots
 - 31.658034197332605587\dots i) \epsilon^2 \nonumber\\
&- (58.38810054205343325\dots
 + 143.03005083114917198\dots i) \epsilon^3 \nonumber\\&- (290.07258795798373896\dots + 95.09318145787845098\dots i) \epsilon^4\\
    \mathrm{m}_{23}\approx& \;(-5.284352448720820476\dots -    23.743525647999454190\dots i) \epsilon^2 \nonumber\\
&-(5.078788429212272033\dots + 
    109.485707877044407942\dots i) \epsilon^3 \nonumber\\
&- (24.814570161293154338\dots +   59.905582552785074450\dots i) \epsilon^4
\end{align*}

\section{Outlook}
\label{sec:outlook}
In this work, we studied one of two planar integral families contributing to $\mu e$ scattering. The integrals were known in the equal mass limit \cite{Henn:2013woa} and in the limit of a vanishing electron mass \cite{Mastrolia:2017pfy} and we were able to generalize these calculation to the case of two different, finite masses. The integrals are expressed in terms of multiple polylogarithms with algebraic function arguments. To our knowledge this is the second time that a differential equation in d-log form with unrationalizable roots in the symbol could be integrated in terms of these functions. We found a surprisingly simple structure of the alphabet after we simplified it using the ideas of Ref. \cite{Heller:2019gkq}. As a next step it would be interesting to see if also the other planar and non-planar integral families contributing to $\mu e$ scattering can be solved in terms of multiple polylogarithms for a finite electron mass.

Another interesting application of the calculation presented in this work could be related to the production of gravitational waves in general relativity. Binary systems producing these waves can be described in perturbation theory in the framework of an effective field theory. At third post-Minkowskian order a two-loop double-box graph, called $H$-graph, that has a similar structure as the diagrams in $\mu e$ scattering contributes. For the equal mass case, the $H$-graph has been recently calculated in Ref. \cite{Kreer:2021sdt}. Having in mind that analytic results for the unequal-mass case in muon-electron scattering are feasible, one can expect that also the $H$-graph with two different masses can be calculated analytically with techniques presented in this paper.

\acknowledgments
The author gratefully acknowledges Andreas von Manteuffel for useful discussions and providing unpublished code that was needed to perform the integration.
The author was supported in part by the German Research Foundation (DFG), through the Collaborative Research Center, Project ID 204404729, SFB 1044, and the Cluster of Excellence PRISMA$^+$, Project ID 39083149, EXC 2118/1.
Our figures were generated using {\tt Jaxodraw} \cite{Binosi:2003yf}, based on {\tt AxoDraw} \cite{Vermaseren:1994je}.
\bibliographystyle{JHEP}
\bibliography{bhabha}

\end{document}